\documentclass[aps,prb,twocolumn]{revtex4}

\usepackage{graphicx}
\usepackage{dcolumn}
\usepackage{amsmath}

\begin{document}

\title{Observation of electromagnon excitations in the BiFe0$_3$ spiral magnet}

\author{M. Cazayous}
\email{maximilien.cazayous@univ-paris-diderot.fr}
\author{Y. Gallais}
\author{A. Sacuto}

\affiliation{Laboratoire Mat\'eriaux et Ph\'enom\`enes Quantiques (UMR 7162 CNRS), 
Universit\'e Paris Diderot-Paris 7, 75205 Paris cedex 13, France\\
Laboratoire de Physique des Solides (UPR5 CNRS),
Ecole Superieure de Physique et de Chimie Industrielle, 10 rue Vauquelin, 
75231 Paris, France}

\author{R. de Sousa}
\affiliation{Departement of Physics and Astronomy, University of Victoria, Victoria, B.C., Canada, V8W 3P6}

\author{D. Lebeugle}
\author{D. Colson}
\affiliation{Service de Physique de l'Etat Condens\'e, DSM/DRECAM/SPEC, CEA Saclay, 91191 Gif-sur-Yvette, France}

\date{\today}

\maketitle

\textbf{Recently, oxide multiferroics have attracted much attention due to their large magnetoelectric effect which allows the tuning of magnetic properties with electric field and vice versa\cite{Eerenstein, Zhao, Senff, Cheong} and open new venues for future spintronic applications such as multiple-state memory devices with dual magnetic and electric control.
BiFeO$_3$ (BFO) belongs to this new class of materials and
shows both ferroelectric and antiferromagnetic orders at room temperature\cite{Smolenski,Ismilzade} with a large electric polarization\cite{Wang, Lebeugle1} associated with a cycloidal spiral magnetic ordering. The incommensurate magnetic order induces magnon zone folding and allows investigations by optical probes of unusual spin waves which couples to optical phonons, the so called "`electromagnons"'. 
Here, we unravel for the first time the electromagnon spectra of BFO by means low energy inelastic light scattering technique. We show the existence of two species of electromagnons corresponding to spin wave excitations in and out of the cycloidal plane. 
The present observations present an unique opportunity to study the interplay between ferroelectric and magnetic orders.}

\par
\par

Low-frequency magneto-optical resonances in the dielectric susceptibility\cite{Pimenov, Mostovoy, Sushkov, Katsura} of multiferroic compounds have attracted much attention the last few years. Optical measurements of magnetic and dynamical properties such as Raman\cite{Singh1, Cazayous, Hermet}, infrared or ellipsometric spectroscopies\cite{Kamba, Lobo} are still in their infancy. In {BFO}, the intimate relationship between electrical polarization and spin wave excitations or magnons is directly related to the strong electromagnetic coupling detected at room temperature in BFO.
A recent theory on spin wave dynamics in incommensurate multiferroic BFO predicted the observation of a series of electromagnon resonances by an optical probe\cite{Sousa}.
\par
Here we report a low-energy Raman scattering study on BFO single crystal where we observe for the first time two species of low-energy electromagnons with two well defined energy level structures. The level structures are linked to the magnon zone folding induced by the periodicity of the cycloid and their coupling with electrical polarization. We assign the two sets of energy levels to the electromagnon mode excitations in and out of the cycloid plane in BFO. 
In addition we show that the temperature dependence of the electromagnon modes drastically contrasts with the one of the optical phonons. The sudden increase of the intensity of the modes and the softening of their frequencies close to T= 140K reveal the occurence of a spin reorientation phase transition in BiFeO$_3$ which confirms the magnetic nature of the electromagnon modes. 

\begin{figure}
\includegraphics*[width=7cm]{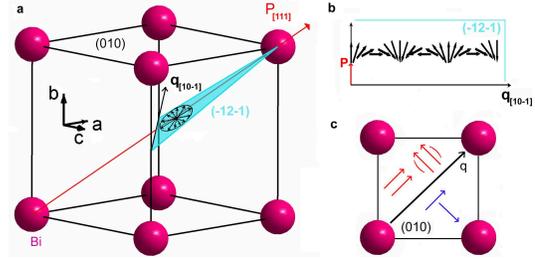}\\
\caption{\label{Fig1} \textbf{Ferroelectric and antiferromagnetic structure of BiFeO$_3$}. \textbf{a}, Structure of the pseudo-cubic BiFeO$_3$ unit cell. Only the Bi atoms are displayed for more clarity. The spontaneous polarisation $P$ is along the [111] direction and the spiral magnet propagation defined by the wavevector $q$ is along the [10-1] direction. \textbf{b}, These two directions define the (-12-1) cycloidal plane (in blue color) where the spin rotation occurs. \textbf{c}, Parallel $(//)$ (red vectors) and crossed polarizations $(\bot)$ (blue vectors) are in the (010) plane and can be defined with respect to the $q$ wavevector. However the exact direction of the [10-1] axis in the (010) plane is not directly accessible and thus the direction of the electrical fields in the parallel polarizations configuration is either perpendicular or parallel to the wavevector $q$.}
\end{figure}

\par
The BiFeO$_3$ single crystals studied were grown in air using a Bi$_2$O$_3$-Fe$_2$O$_3$ 
flux technique\cite{Lebeugle}. They have a millimeter size and have the shape of a chrismas tree with a large surface plane corresponding to the (010) face in the pseudo-cubic representation. The spontaneous electrical polarization has a magnitude between 50 and 100 $\mu$C/cm$^2$ along [111] direction\cite{Lebeugle} and makes an angle of 54 degrees wih the (010) face (Fig.\ref{Fig1}a).
 It comes from a rhombohedrally distorted perovskite structure below a high Curie temperature ($T_c$$\sim$1100 K) above which the  BFO structure is pseudo-cubic. From a magnetic point of view, BFO exhibits a G-type antiferromagnetic order below the N\'eel temperature ($T_N$$\sim$640 K)
\cite{Smolenski,Ismilzade}. The magnetic order is subjected to long range modulation associated with a cycloidal spiral with a period length of approximatively 62~nm\cite{Sosnowska}. The spiral propagation is along [10-1] direction with a spin rotation within (-12-1) plane (Fig.\ref{Fig1}a and b).

\par
Raman measurements have been performed 
using the 647.09 nm (1.9 eV) excitation 
line from a Ar$^+$-Kr$^+$ mixed gas laser. 
The Raman scattering spectra were recorded between  
7 and 300~K using a triple spectrometer Jobin Yvon T64000 with the high rejection rate in order 
to measure at the lowest frequencies.

\begin{figure}
\includegraphics*[width=7cm]{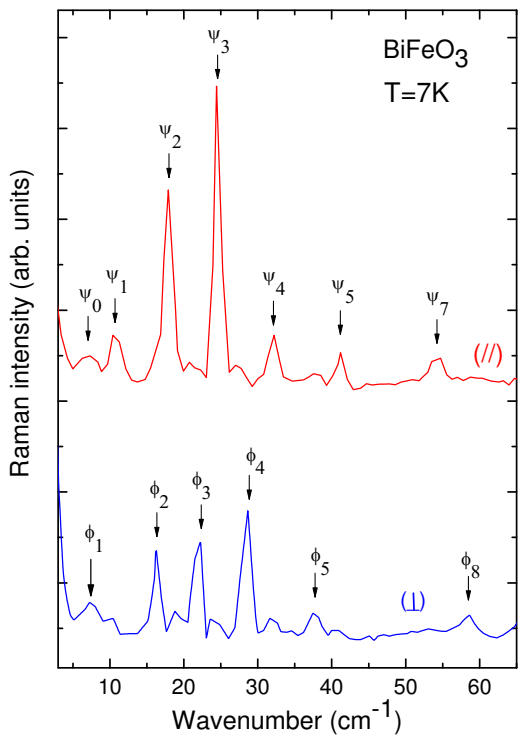}\\
\caption{\label{Fig2} \textbf{Raman spectra of electromagnons in BiFeO$_3$}. The equally and non equally spaced modes at low frequencies correspond respectively to $\Psi$ and $\Phi$ electromagnon modes selected using parallel $(//)$ and crossed $(\bot)$ polarizations as defined in Fig.~\ref{Fig1}.}
\end{figure}

\par
\begin{figure}
\includegraphics*[width=7cm]{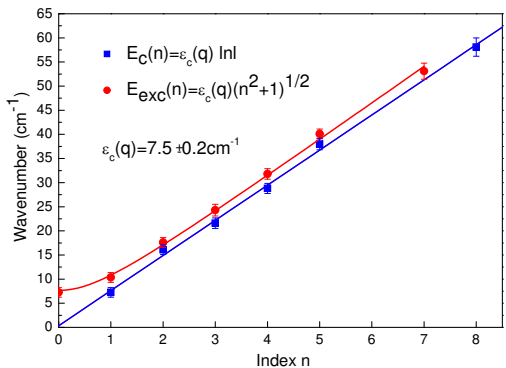}\\
\caption{\label{Fig3} \textbf{Energy levels of electromagnons}. $\Phi$ (blue squares) and $\Psi$ (red full circles) mode frequencies as a function of the mode index $n$. The expressions of the energy levels fit well the experimental data with a cyclon energy of $\epsilon_c(q)$=7.5$\pm$0.2cm$^{-1}$.}
\end{figure}

Figure~\ref{Fig2} shows the low frequency Raman spectra of BFO single crystal. The incoming and outgoing light polarizations are in the (010) pseudo cubic plane which contains the cycloidal propagation wave vector. Although no phonons are expected under 50 cm$^{-1}$,\cite{Kamba, Haumont} several peaks can be observed. 
Twelve peaks are detected with the first lying at 7.4$\pm$0.5~cm$^{-1}$ very closed to the stray-light centered at zero energy and the last one at 59$\pm$0.5~cm$^{-1}$. By choosing parallel $(//)$ polarizations (see Fig.~\ref{Fig1}c) and crossed polarizations $(\bot)$ in the (010) plane we have successfully selected two distinct sets of peaks. They are respectively denoted $\Psi$ and $\Phi$ and their energy is reported in Fig.~\ref{Fig2}. A first glance on Fig.~\ref{Fig2} reveals that the peak locations of the $\Phi$ modes are equally spaced starting at zero frequency and remarkably correspond to an arithmetical sequence of ratio close to  $7.4~cm^{-1}$ (the sixth and seventh modes are missing). 
By contrast the sequence of the $\Psi$ modes is not regularly spaced at low frequencies as it can be seen in Fig.~\ref{Fig3} where the
$\Phi$ and $\Psi$ modes frequencies are plotted as a function of the $n$ mode index. The $n$ index labels the modes from their lowest to highest energy. 

\par
To go further, we have interpreted our Raman spectra based on a recent theoretical work which predicts the possible observation of electromagnons in incommensurate BiFeO$_3$ spiral magnet by optical spectroscopy \cite{Sousa}. This approach is based on the effective Ginzburg-Landau free energy which includes the coupling between the spin waves and the electrical polarization. This approach leads to two species of electromagnon excitations which lie in and out of the cycloidal plane, respectively the cyclon and the extra-cyclon modes, with distinctive dispersive energy curves that depend on their coupling to the electrical polarization. Crucially, the two sets of hybridized spin wave modes are expected to be detected by optical spectroscopies at small wavevectors ($k\approx 2\pi/ 5000 \AA~$) such as Raman scattering thanks to the zone folding of the antiferromagnetic (AF) magnons induced by the incommensurabilty of the spiral magnet. 
The AF magnon zone folding leads to a very simple expression for the energy level structure of the cyclon mode which remains gapless as expected from the AF ordering. The extra-cyclon energy level structure however, is expected to acquire a gap due to the pinning of the cycloidal plane by the electrical polarization in full agreement with our observation \cite{Sousa} (see Fig. \ref{Fig3}). 
The energy levels sequences are respectively given by  $E_c(n)=\epsilon_c(q)\left|{n}\right|$ and $E_{exc}(n)=\epsilon_c(q)\sqrt{n^2+1}$ . $\epsilon_c(q)$ is the cyclon energy at the wavevector $q$ which is simply related to the AF magnon velocity $v$: $\epsilon_c(q)=\sqrt{v}q$. 
In this picture only one adjusting parameter, i.e.  the cyclon energy (or equivalently the AF magnon velocity) is needed. It is therefore straight-forward to compare our experimental data to the theory  by plotting the $\Phi$ and $\Psi$ modes frequencies as a function of the mode index $n$ and fitting the data with respect to the cyclon and extra-cyclon energy levels equations.

\par
The results are shown in Fig.~\ref{Fig3}.  Our experimental data remarquably fit to the equations and we can unambiguously assigned the $\Phi$ sequence to the energy levels of the cylon modes  and the $\Psi$ sequence to the extra cyclon modes. 
The best agreement is obtained for a cyclon energy $\epsilon_c(q)$ equal to 7.5$\pm$0.2~cm$^{-1}$. 
The estimated cyclon energy $\epsilon_c(q)$ calculated by de Sousa {\it et al.} is about 10$^{12}$ rad/s equivalent to 5.3~cm$^{-1}$ and calculated in order to match the susceptibility measured on thin films. This is close to our experimental findings performed in bulk single cristal. The magnon velocity $v$ can be estimated using the magnon wavevector $q=2\pi / 62~nm~$ which gives a velocity of 1.4.10$^6$~cm.s$^{-1}$.

\begin{figure}
\includegraphics*[width=9cm]{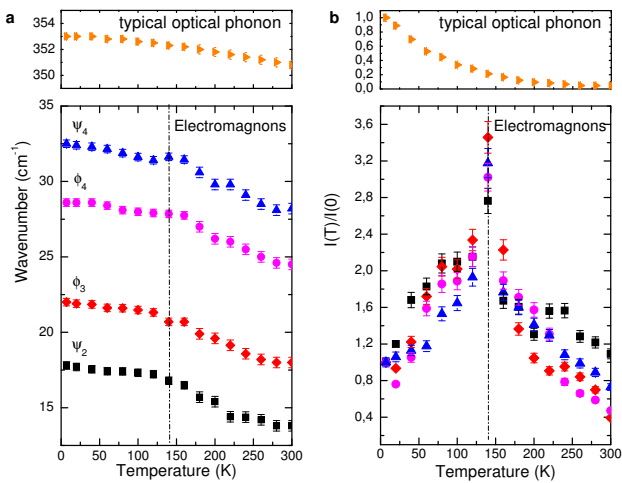}\\
\caption{\label{Fig4} \textbf{Magnetic signature of the electromagnons}. \textbf{a}, Temperature dependance 
of the four electromagnon energies ($\Psi_2$, $\Phi_2$, $\Phi_4$ and $\Psi_4$) 
and of one typical optical phonon mode. The decreasing slope of the electromagnon wavenumber presents a anomaly around 140~K. 
\textbf{b}, Integrated intensity (area under the Raman peaks) of the electromagnon and phonon excitation. The electromagnon intensity exhibits a maximum at 140~K unlike the phonon mode.}
\end{figure}

In Fig.~\ref{Fig4}a is shown the evolution of the electromagnon energy levels as a function of temperature. The typical variation of an optical phonon is also reported in Fig.~\ref{Fig4}a.  We can clearly see that the temperature dependence of the electromagnon energies reveals an anomaly close to T=140~K whereas the optical phonon energy continuously decreases as a function of temperature. Interestingly the Raman integrated intensities of the electromagnon peaks exhibits a maximum at T=140~K as shown in Fig.~\ref{Fig4}b while the integrated intensity of the phonons remains featureless: the integrated intensity of a typical optical phonon is shown in Fig.~\ref{Fig4}b and slowly decreases as the temperature is raised with no abrupt change at 140~K. 

Similar softening has already been observed in the magnon temperature dependence of rare-earth 
orthoferrites RFeO$_3$ (R being a rare-earth atom)\cite{Koshizuka, Venugopalan} and assigned to spin reorientation phase transition of the ordered Fe$^{3+}$ magnetic moments. At room temperature, the Fe$^{3+}$ magnetic moments in BiFeO$_3$ is confined in (-12-1) cycloidal plane (Fig.~\ref{Fig1}a). Our measurements suggest a small spin reorientation out of this plane.

\par
Our study has revealed two species of electromagnons in BiFeO$_3$: 
the cyclon and extra-cyclon. They present two well defined sequences of energy levels at $k=0$: the cyclon sequence remains gapless while the extra-cyclon sequence is gapped by the pinning of the cycloid plane by the electrical polarization. 
The present experimental findings are made possible by the incommensurability of the spiral magnetic ordering of BFO and open new venues to study electromagnon excitations under magnetic and electric fields via optical probes. We hope they will reveal the intimate relationship between magnetic and ferroelectric magnetic orders in multiferroic materials. 

\par
Acknowledgements
\par
The authors would like to thank R. Lobo and Ph. Monod for helpful 
discussions and for a critical reading of the manuscript.

\end{document}